\documentclass[aps, prd, twocolumn, superscriptaddress, preprintnumbers, longbibliography,nofootinbib]{revtex4-1}

\usepackage[top=2.8cm, bottom=2.8cm, left=2cm, right=2cm]{geometry}
\usepackage[utf8]{inputenc} 
\usepackage[T1]{fontenc}
\usepackage{amsmath, amssymb, lmodern} 
\usepackage[caption=false]{subfig}
\usepackage{graphicx}
\usepackage{natbib}
\usepackage{booktabs}
\usepackage{color}
\usepackage{hyperref}
\hypersetup{colorlinks, citecolor = blue, filecolor = blue, linkcolor = blue, urlcolor = blue}

\definecolor{darkgreen}{rgb}{0,0.6,0}

\begin{document}

\preprint{IFT-UAM/CSIC-22-127}

\title{Using machine learning to compress the matter transfer function $T(k)$}

\author{J. Bayron Orjuela-Quintana}
\email{john.orjuela@correounivalle.edu.co}
\affiliation{Departamento  de  F\'isica,  Universidad  del Valle, Ciudad  Universitaria Mel\'endez,  Santiago de Cali  760032,  Colombia}
\affiliation{Instituto  de  F\'isica  Te\'orica  UAM-CSIC,  Universidad  Auton\'oma  de  Madrid,  Cantoblanco,  28049  Madrid,  Spain}

\author{Savvas Nesseris}
\email{savvas.nesseris@csic.es}
\affiliation{Instituto  de  F\'isica  Te\'orica  UAM-CSIC,  Universidad  Auton\'oma  de  Madrid,  Cantoblanco,  28049  Madrid,  Spain}

\author{Wilmar Cardona}
\email{wilmar.cardona@unesp.br}
\affiliation{ICTP South American Institute for Fundamental Research \& Instituto de F\'isica Te\'orica, Universidade Estadual Paulista, 01140-070, S\~ao Paulo, Brazil}

\begin{abstract}
The linear matter power spectrum $P(k,z)$ connects theory with large scale structure observations in cosmology. Its scale dependence is entirely encoded in the matter transfer function $T(k)$, which can be computed numerically by Boltzmann solvers, and can also be computed semi-analytically by using fitting functions such as the well-known Bardeen-Bond-Kaiser-Szalay (BBKS) and Eisenstein-Hu (EH) formulae. However, both the BBKS and EH formulae have some significant drawbacks. On the one hand, although BBKS is a simple expression, it is only accurate up to $10\%$, which is well above the $1\%$ precision goal of forthcoming surveys. On the other hand, while EH is as accurate as required by upcoming experiments, it is a rather long and complicated expression. Here, we use the Genetic Algorithms (GAs), a particular machine learning technique, to derive simple and accurate fitting formulae for the transfer function $T(k)$. When the effects of massive neutrinos are also considered, our expression slightly improves over the EH formula, while being notably shorter in comparison.
\end{abstract}

\maketitle

\section{Introduction} 
\label{Sec: Introduction}

Almost two decades ago, the $\Lambda$CDM model was designated as the standard model of cosmology \cite{Peebles2020}. The concordance model is supported by several observations including the angular power spectrum of the cosmic microwave background \cite{Planck:2018vyg}, the distribution of galaxies at large scales \cite{DES:2017qwj, Abbott:2018xao,DES:2021wwk}, the late-time accelerated expansion of the Universe \cite{Riess:1998cb, perlmutter:1998np, SupernovaSearchTeam:2004lze}, the acoustic peaks as a fingerprint of the baryon acoustic oscillations in the early Universe \cite{Percival_2010,  Blake_2011, Aubourg:2014yra}. These are just some of the many observational tests the $\Lambda$CDM model has overcome \cite{deBernardis:2000sbo, Jaffe:2003it, Planck:2018jri, Planck:2019evm,DES:2018ekb}. 

However, it is fair to mention that there exist some discrepancies in $\Lambda$CDM, such as the $H_0$ and $\sigma_8$ tensions \cite{Riess:2019cxk, Abdalla:2022yfr}. In general, alleviation of any of these tensions and others (see Refs. \cite{Joyce:2016vqv, diValentino:2021izs, Heisenberg:2022lob, Abdalla:2022yfr, Perivolaropoulos:2021jda,Cardona:2022pwm}) requires either the introduction of non-standard matter fields \cite{guo:2018ans, Agrawal:2019lmo, Heisenberg:2020xak}, or modifications to gravity  \cite{Clifton:2011jh, Heisenberg:2018mxx, Cardona:2022lcz}. Nonetheless, these theoretical alternatives, although physically viable in light of observations, usually introduce additional parameters, and hence they are not preferred over $\Lambda$CDM which is described by only six-parameters and provides an excellent fit to most of available data. In summary, $\Lambda$CDM is the simplest and most accurate model we currently have, given that no extensions to this paradigm have been detected so far \cite{heavens:2017hkr, singh:2021mxo, arjona:2021mzf, Euclid:2022ucc, Cardona:2020ama}.

One of the most important cosmological probes we have in favor of the concordance model is the  distribution of galaxies at large scales \cite{SDSS:2003tbn, BOSS:2016wmc}. In order to contrast our theoretical predictions against these observational data, it is necessary to extract the statistical information in the distribution of the large scale structure by computing the matter power spectrum, $P(k, z)$, which depends on the scale $k$ and the redshift $z$. It can be shown that at first order in cosmological perturbations, and neglecting neutrinos, the dependence of $P(k, z)$ on $k$ is encoded in the so-called matter transfer function $T(k)$, while the dependence on $z$ is encoded in the growth factor $D_+ (z)$ \cite{dodelson2020modern}. Therefore, for a fixed redshift, the matter power spectrum is a function only of the scale and its form is mostly described by the matter transfer function (see Sec. \ref{Sec: The Matter Transfer Function}).

The calculation of $P(k, z)$ generally requires to solve the multi-species Boltzmann equations, which can be done numerically in a matter of seconds using Boltzmann solvers, like the codes \texttt{CLASS} \cite{Blas_2011} and \texttt{CAMB} \cite{Lewis:1999bs}. However, to have an accurate analytical description for specific quantities is always desirable.  Following this line of thought, Bardeen, Bond, Kaiser, and  Szalay found a fitting function for the transfer function considering radiation, baryons, and cold dark matter \cite{Bardeen:1985tr}. This BBKS formula is accurate up to $10\%$ which is well-below the precision of current data \cite{Turner:2022gvw}. A better alternative is the fitting formula given by Eisenstein and Hu (EH) in Ref. \cite{Eisenstein:1997ik}. The EH formula achieves a precision of around $1$-$2\%$, however it is given in terms of around 30 different, complicated expressions. These fitting formulae have been extensively used in the literature \cite{Boyanovsky:2008he, Dvornik:2022xap, Schoneberg:2022ggi}.

Machine learning has long been exploited in physics (see Ref.~\cite{carleo:2019ptp} for a review). In particular, machine learning has been used to address symbolic regression problems, i.e., finding an analytical expression that accurately describe a given data set \cite{schmidt:2019doi, brunton:2016dac, dcGP, Udrescu:2019mnk, cranmer:2020wew, Liu:2021azq}. In this work, we use a specific machine learning technique known as Genetic Algorithms (GAs). GAs are loosely based on biological evolution concepts. In a nutshell, they attempt to improve the goodness of fit of the candidate expressions by randomly combining them and/or modifying some parts of them \cite{koza1992genetic}. This approach seems to be suitable for finding analytical formulae for quantities of interest in cosmology \cite{Arjona:2020kco, EUCLID:2020syl, Arjona:2021hmg, Euclid:2021cfn, Aizpuru:2021vhd, Euclid:2021frk, Alestas:2022gcg}. Here, we use GAs to find analytical functions for the matter transfer function with accuracy of around 1\% while being significantly shorter and thus easier to handle than other available formulae.

The organization of this paper is as follows. In Sec. \ref{Sec: Previous Fitting Formulae}, we discuss existing fitting functions for the matter transfer function, while some generalities on the GA are given in Sec.~\ref{Sec: Genetic Algorithm}. Then, we present our results in Sec.~\ref{Sec: Results}, and in Sec.~\ref{Sec: Conclusions} we summarize our conclusions.

\section{The Matter Transfer Function}
\label{Sec: The Matter Transfer Function}

As mentioned in Sec.~\ref{Sec: Introduction}, on large scales where non-linearities are negligible, the linear  matter power spectrum can be compared against observations of galaxy clustering, and gravitational lensing, among others \cite{Reid_2010, DiazRivero:2018oxk, Chabanier:2019eai}. 
In the concordance model, the primordial curvature perturbations generated during inflation are related to the gravitational potential at late times by means of two functions: $i)$ the matter transfer function $T(k)$ which encodes its scale dependence and describes the evolution of perturbations during horizon crossing and transition from radiation to matter domination, $ii)$ the growth factor $D_+(a)$ which describes the time-dependent growth of 
matter density perturbations at late times. Thus, the gravitational potential can be written as
\begin{equation}
\Phi(k, a) \propto \mathcal{R}(k) T(k) D_+ (a),  
\end{equation}
where $\mathcal{R}$ is the primordial curvature perturbation, and $a$ is the scale factor. At late times, for sub-horizon modes, and neglecting massive neutrinos, this gravitational potential can be related to the matter contrast $\delta_m$ by means of the Poisson equation:
\begin{equation}
\label{Eq: Phi(k, a)}
k^2 \Phi(k, a) \propto \rho_m a^2 \delta_m (k, a),
\end{equation}
where $\rho_m$ is the background density of pressure-less matter. Taking into account that the primordial perturbations are nearly Gaussian with zero mean  \cite{Planck:2018jri}, the linear matter power spectrum can be written as \cite{dodelson2020modern} 
\begin{equation}
\label{Eq: P(k, a)}
P(k, a) \propto \frac{k^{n_s}}{\Omega_m^2} D_+^2 (a) T^2 (k), 
\end{equation}
where $n_s$ is the scalar spectral index of primordial fluctuations, and $\Omega_m$ is the density parameter of pressure-less matter. Therefore, for a fixed redshift, the power spectrum is given by
\begin{equation} \label{Eq: P(k)}
P(k) \propto k^{n_s} T^2 (k),
\end{equation}
namely, it is fully determined by the transfer function. In this work, we present a very accurate fitting formula for $T(k)$ as a function of the density of baryons and matter, such that it is straightforward to compute the linear matter power spectrum. 

As it is well-known from experiments, neutrinos are massive \cite{, Lesgourgues:2013sjj, KATRIN:2021uub}. At sufficiently small scales, free streaming massive neutrinos imprint their effects on the cosmological evolution, which translates to a further suppression of the matter power spectrum. In this case, the growth factor acquires a scale dependency, making non-trivial a similar separation as in Eq.~\eqref{Eq: P(k, a)}. However, for a fixed redshift, it is possible to absorb all the scale dependent effects in an effective matter transfer function, as shown in Ref.~\cite{Eisenstein:1997jh}.

\section{Previous Fitting Formulae}
\label{Sec: Previous Fitting Formulae}

Before the advent of fast and accurate Boltzmann solvers, several attempts were made to describe analytically the matter transfer function. One of the most remarkable results of this pursuit is the BBKS formula which is based on previous fitting formulae by Bardeen, Bond, Efstathiou, and Szalay \cite{83ApJ...274..443B, 84ApJ...285L..45B, Kolb:1990vq}. In the case that $\Omega_b \ll \Omega_m$, that is, the main contribution to the matter content is in the form of Cold Dark Matter (CDM), the BBKS formula reads
\begin{align}
T_{c, \text{BBKS}} (k) &\equiv \frac{\ln (1 + 2.34 q)}{2.34 q} \Big[ 1 + 3.89 q  \nonumber \\
 &+ ( 16.1 q )^2 + ( 5.46 q )^3 + ( 6.71 q )^4 \Big]^{-1/4},
\end{align}
where
\begin{equation}
q(k) \equiv \frac{k \theta^{1/2}}{(\omega_m - \omega_b) \text{Mpc}^{-1}}, \qquad \theta \equiv \frac{\rho_r}{1.68 \rho_\gamma}, 
\end{equation}
and $\omega_X \equiv \Omega_X h^2$ the reduced density parameter ($\Omega_X$ being the density parameter of species $X$), $h$ the reduced Hubble constant, $\rho_X$ the background density, and $X = b$, $c$, $m$, $r$, $\nu$, $\gamma$ denotes baryons, CDM, pressure-less matter, radiation, neutrinos, photons, respectively. When $\omega_b$ is not negligible, the transfer function is modified as
\begin{equation}
\label{Eq: BBKS}
T_\text{BBKS}(k; \omega_b, \omega_m) \equiv T_{c, \text{BBKS}} (k) \left[ 1 + \frac{\left( k R_\text{Jr} \right)^2}{2} \right]^{-1},
\end{equation}
where
$R_\text{Jr} \equiv 1.6 ( \omega_m - \omega_b )^{-1/2} \, \text{kpc}$.

A more accurate formula was presented by Eisenstein and Hu in Ref.~\cite{Eisenstein:1997ik}. This formula is constructed from several physically motivated terms, such as acoustic oscillations, Compton drag, velocity overshoot, baryon in-fall, adiabatic damping, Silk damping, CDM suppression. These terms accurately describe, for instance, the  suppression in the transfer function on small scales due to the presence of baryons, and the amplitude and location of baryonic oscillations \cite{Eisenstein:1997ik}. All the mentioned effects are taken into account in the EH formula by around 30 expressions, which we present in Appendix~\ref{App: EH Formula}. 

The free-streaming scale of massive neutrinos is imprinted in the transfer function, which translates to a further suppression on the smallest scales. In Ref.~\cite{Eisenstein:1997jh}, Eisenstein and Hu provided a fitting formula when baryons, CDM and massive neutrinos are considered. The latter formulation also requires around 30 expressions, which we write in Appendix~\ref{App: EH Formula Neutrinos}. 

Although accurate and based on known physics, the EH expressions are, at the end, fitting formulae and not a fundamental result. As we will show, our GA fitting formulae are notably shorter, as accurate as the EH expression when only baryons and CDM are considered, and slightly more accurate when the effects of massive neutrinos are non-negligible. Therefore, our GA provides  compelling alternatives to other available fitting formulae.

\section{Genetic Algorithms}
\label{Sec: Genetic Algorithm}

Here we briefly review the GA, which can be used as a stochastic symbolic regression approach. Basically, the symbolic regression problem consists in finding a fitting function for a given dataset, and thus defining a non-parametric approach to describe data. There are several free and commercial codes to perform symbolic regression \cite{schmidt:2019doi, brunton:2016dac, dcGP, Udrescu:2019mnk, cranmer:2020wew, Liu:2021azq}. The GAs have been widely used in several branches of physics, like particle physics \cite{Allanach:2004my, Akrami:2009hp, Abel:2018ekz}, astrophysics \cite{Luo:2019qbk, ho:2019zap}, and cosmology \cite{Bogdanos:2009ib, Nesseris:2010ep, Nesseris:2012tt, Nesseris:2013bia, Sapone:2014nna, Arjona:2019fwb}.

In the GA, which is a machine learning technique inspired by concepts in evolutionary biology \cite{koza1992genetic}, a population composed by several individuals (mathematical expressions in this case) evolves expecting to optimize the goodness of fit of next generations to a given dataset, which can be measured by  a goodness of fit function, usually taken to be a $\chi^2$. Each individual is in turn composed by a selected set of basic operations, i.e., the grammar, which are randomly selected depending on the ``nucleotides''. These nucleotides are random numbers which in group form the genes setting the mathematical expression of an individual. Once the initial population is created, i.e., the progenitors, the goodness of fit of each individual is measured and, using a tournament scheme, a set containing the best prospects is selected for reproduction and survival to contribute to the next generation. 

The reproduction is carried out by the so-called genetic operators, namely, crossover and mutation. In the crossover stage, two individuals are randomly combined to produce new individuals sharing characteristics of both parents. In the mutation stage, one nucleotide of an individual is randomly selected and modified.  This process is repeated for a given number of generations.

\begin{figure}[t!]
\includegraphics[width = 0.38\textwidth]{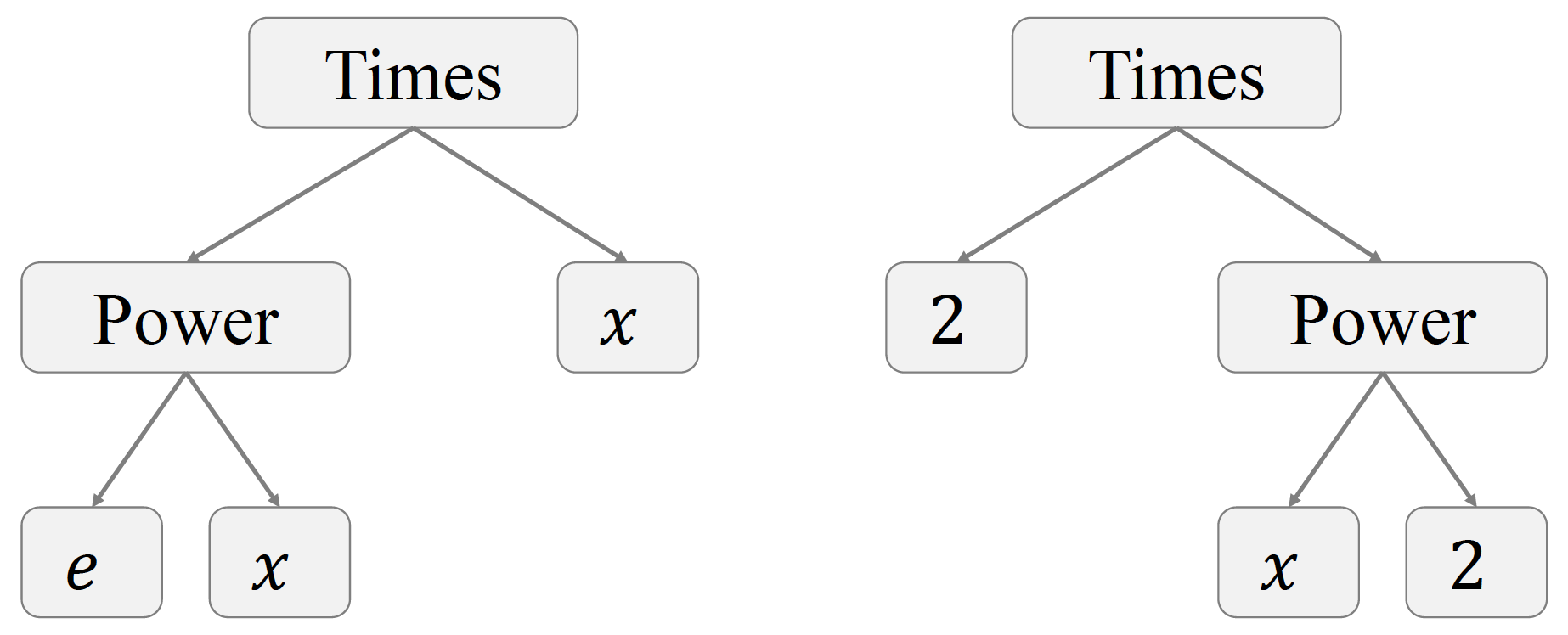}
\caption{Two exemplifying grammatical expressions corresponding to the individuals $x e^x$ and $2 x^2$.}
\label{Plot: Individuals}
\end{figure}

\begin{figure}[t!]
\includegraphics[width = 0.3\textwidth]{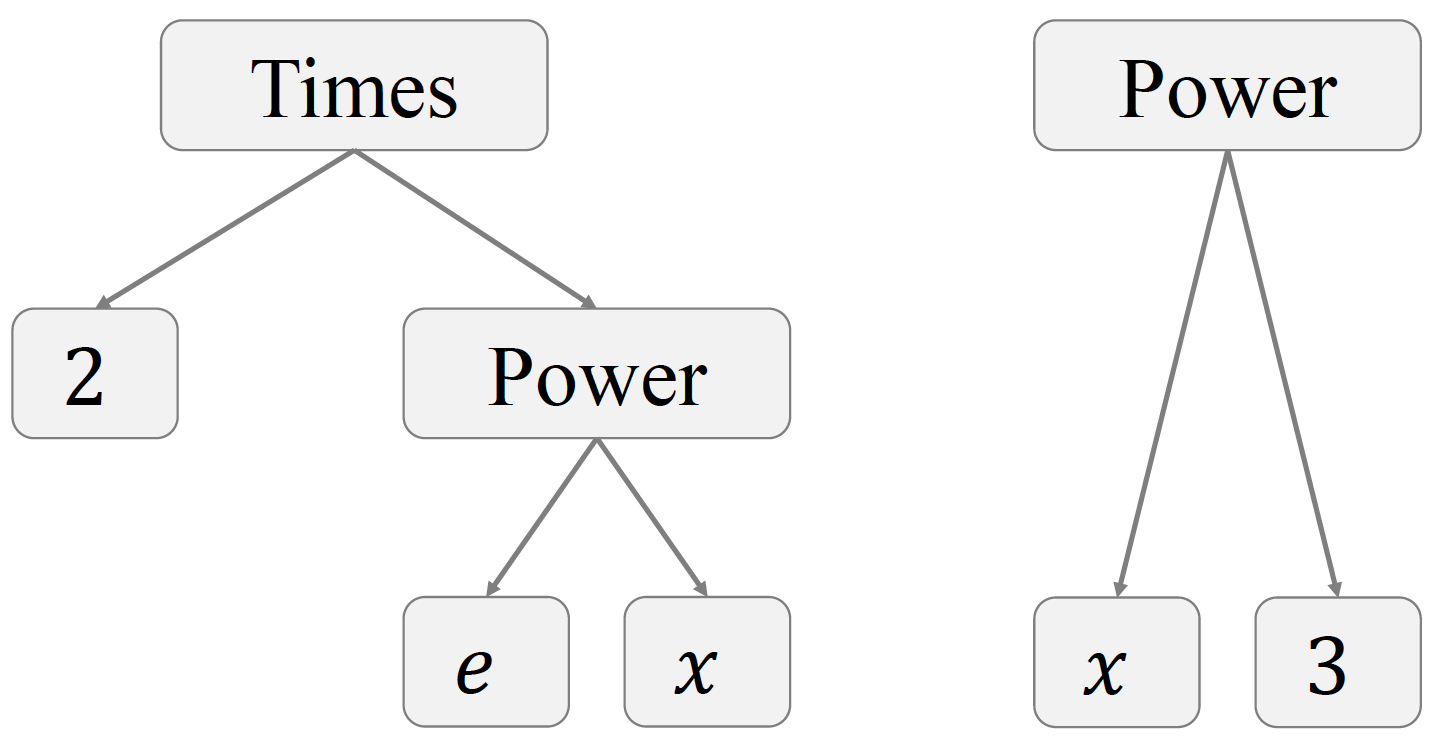}
\caption{The two selected individuals have been combined to produce two new individuals: $2 e^x$ and $x^3$.}
\label{Plot: Crossover}
\end{figure}

\begin{figure}[t!]
\includegraphics[width = 0.32\textwidth]{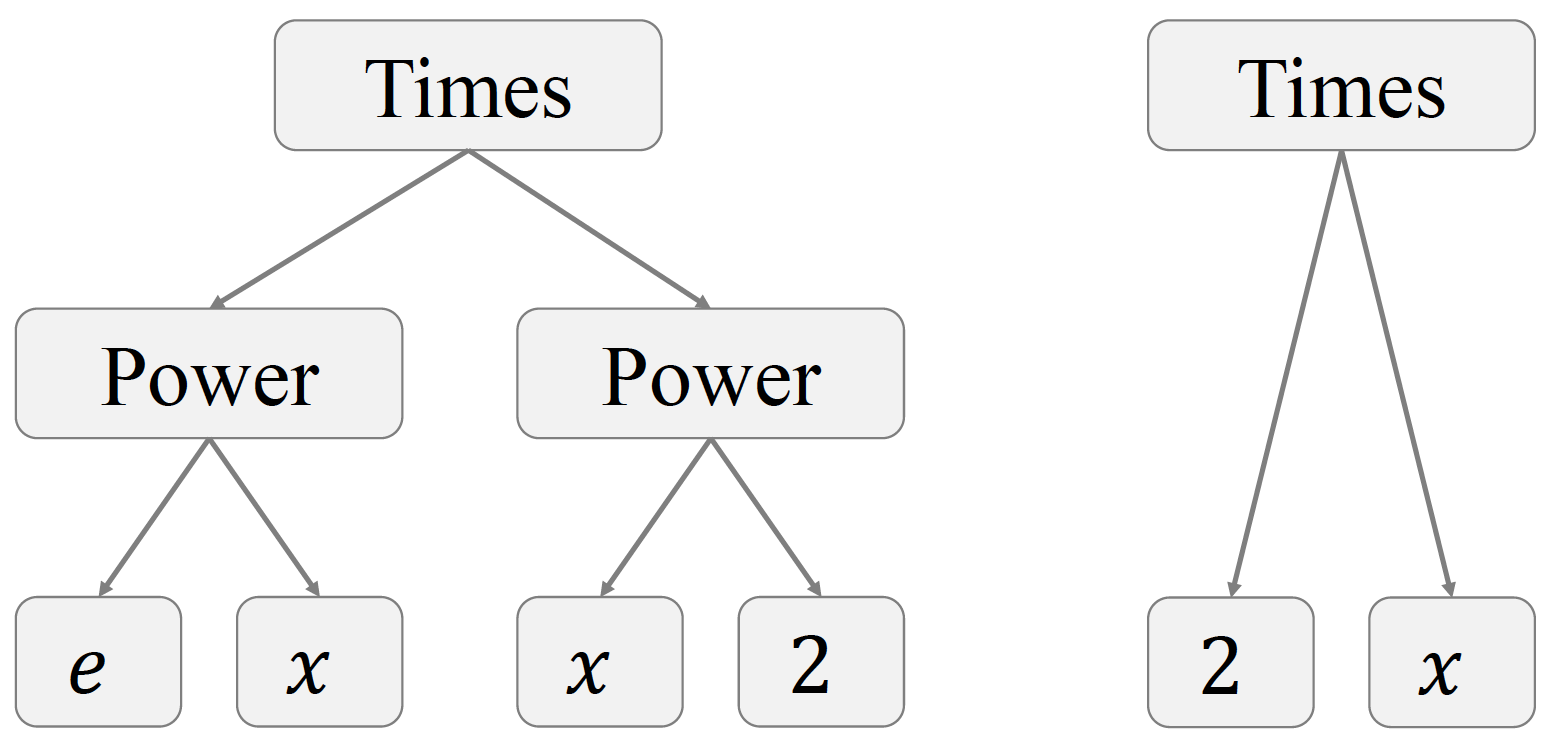}
\caption{The two selected individuals have suffered a mutation to become two new individuals: $x^2 e^x$ and $2 x$.}
\label{Plot: Mutation}
\end{figure}

Now, we describe the way reproduction works in GA with an example. In Fig. \ref{Plot: Individuals}, there are two expressions in tree representation, i.e., two individuals $x e^x$ and $2 x^2$. Let us assume that these individuals were selected for reproduction. Figure \ref{Plot: Crossover} shows possible combinations of the two individuals, while Fig. \ref{Plot: Mutation} shows their possible mutations. For instance, the new individual in the left-hand side of Fig. \ref{Plot: Crossover} is born from the combination of the expressions $e^x$ and $2$ from its parents, while the other individual is made from the remaining parts, $x$ and $x^2$, using the proper basic operation ``Times'' or product. During the mutation stage, a selected individual is altered randomly. As shown in Fig. \ref{Plot: Mutation}, for instance, the individual $x e^x$ mutated to $x^2 e^x$, and the individual $2x^2$ mutated to $2 x$.

\section{Results}
\label{Sec: Results}

In this section, we present our fitting formulae for the matter transfer function. We firstly consider the case when $T(k)$ depends only on the amount of baryons and matter, and then we add the effects of massive neutrinos. For both cases, we start describing how the data for the fitting process was gathered using \texttt{CLASS}, and later we proceed to introduce the simple fitting function we found using the GA.

\subsection{Baryons and Matter}
\label{Sec: Baryons and Matter}

\subsubsection{\textbf{Data}}

We use the Boltzmann solver \texttt{CLASS} to compute the linear matter transfer function\footnote{Note that \texttt{CLASS} can take into account non-linear effects via fitting functions such as HALOFIT \cite{Smith:2002dz}. Here however we focus on the linear regime and neglect non-linear contributions when generating the data.} $T(k)$.  In this code, each pair of parameters $\{ \omega_b, \omega_m \}$ defines a cosmology for which we can compute the gravitational potential $\Phi$ as a function of the scale $k$ at a fixed redshift. Hence, to see the dependence of the transfer function on these parameters, we make a grid of $4\times4$ pairs of $\{ \omega_b, \omega_m \}$ and compute the gravitational potential for each pair. We consider that $\omega_b \in [ 0.0214, 0.0234 ]$, and $\omega_m \in [0.13, 0.15]$, which are around $10\sigma$ from the best-fit values found by the Planck Collaboration \cite{Planck:2018vyg}. For each considered cosmology (16 in total), \texttt{CLASS} retrieves $114$ points $\{k, \Phi \}$. Therefore, our preliminary dataset is composed by 1824 points whose rows are given as $\{k, \omega_b, \omega_m, \Phi \}$. Now, since the transfer function is basically the normalized potential, we normalize each of the 16 sub-datasets to get $T(k; \omega_b, \omega_m)$. At the end, our final dataset is a table of dimensions $1824\times4$ with data points given as $\{k, \omega_b, \omega_m, T \}$.  

\begin{table}[t!]
\setlength{\tabcolsep}{12pt}
\begin{tabular}{ccc}
\toprule 
 & Expression & $\%$ Accuracy \tabularnewline
\midrule
\midrule 
BBKS & Eq. \eqref{Eq: BBKS} & 8.70957 \tabularnewline
EH & Appendix~\ref{App: EH Formula} & 0.777504 \tabularnewline
GA & Eq. \eqref{Eq: TGA} & 0.815012 
\tabularnewline
\bottomrule
\end{tabular}
\caption{Expression and accuracy of the fitting formulae for the matter transfer function as a function of $k$, $\omega_b$, and $\omega_m$. Although the GA does not provide a more accurate formula than the EH formula, the simplicity of the former [Eq.~\eqref{Eq: TGA}] is a major improvement over the EH formula (Appendix~\ref{App: EH Formula}).}
\label{Table: Accuracy}
\end{table}

We quantify the goodness of fit of a given analytical expression by the following function 
\begin{equation}
\label{Eq: chi2 GA}
\text{\%Acc} = \frac{100}{N}\sum_{i = 1}^N \left| \frac{T_{i, \text{CLASS}}- T_{i, \text{analytical}}}{T_{i, \text{CLASS}}} \right|,
\end{equation}
where $N = 1824$ is the number of data points in our dataset, and $T_i$ is a simplified notation for the transfer function evaluated at $k_i$, $\omega_{b_i}$, and $\omega_{m_i}$. 

\subsubsection{\textbf{Fitting Formula from the GA}}

Here, we give a few details concerning the specific GA configuration we used and present our fitting formula.

As explained in Sec.~\ref{Sec: Genetic Algorithm}, the GA looks for a fitting formula to a dataset by evolving the population, which is composed by expressions combining specific operations (grammar) and coefficients. Schematically, our GA is searching for a function of the form
\begin{equation}
\label{Eq: TGA 1}
T_\text{GA}(x) \equiv \left[ 1 + f_\text{GA}(x) \right]^{-1/4}, 
\end{equation}
where
\begin{equation}
\label{Eq: fGA}
x \equiv \frac{k \, \text{Mpc}}{\omega_m - \omega_b}, \quad f_\text{GA} \equiv \sum_{i = 1}^{4} a_i g_i \left( (b_i x
)^{c_i} \right).
\end{equation}
We see that $x$ is a dimensionless quantity since $k$ is given in units $[h \, \text{Mpc}^{-1}]$, $h$ the reduced Hubble parameter. The constants $a_i$, $b_i$, $c_i$ are non-negative random numbers, and $g_i$ is an operation in the grammar set, which we consider to be simply of the form $\{ x \}$. The quantities $a_i$, $g_i$, $b_i$, and $c_i$ are the 4 nucleotides. The sum goes from 1 to 4 because we are considering that the genome of the individuals is formed by 4 genes, each one composed by the 4 nucleotides, such that the chromosomes are $4\times4$ matrices. 

It should be noted that the number of genes has to be fixed before the code starts running and we have chosen the number four for two reasons: i) first, from experience we have found that a number of four genes is a compromise (found by trial and error) on the length of the function, as less than is not enough to fit the data, while a much larger number slows down the code considerably,
ii) second, if the number of genes is too high, then this could obviously lead to overfitting or instabilities in the fitting. Furthermore, even though at first sight it seems so, the parameters $a_i$ and $b_i$, in general are not degenerate as $b_i$ appears inside the grammar function, however there may be some degeneracies only when the grammar is linear.

Overall, this configuration, although restrictive, fixes the length of the expressions, thus avoiding over-fitting problems. A couple of reasons motivate  the specific forms of Eqs.~\eqref{Eq: TGA 1} and \eqref{Eq: fGA}: $i)$ the transfer function has specific limits: $T \rightarrow 0$ when $k \rightarrow \infty$, and $T \rightarrow 1$ when $k \rightarrow 0$, $ii)$ the transfer function only takes on non-negative values, and $iii)$ $T(k)$ is smooth. 

\begin{figure*}
\centering
\begin{minipage}[b]{.45\textwidth}
\includegraphics[width=\textwidth]{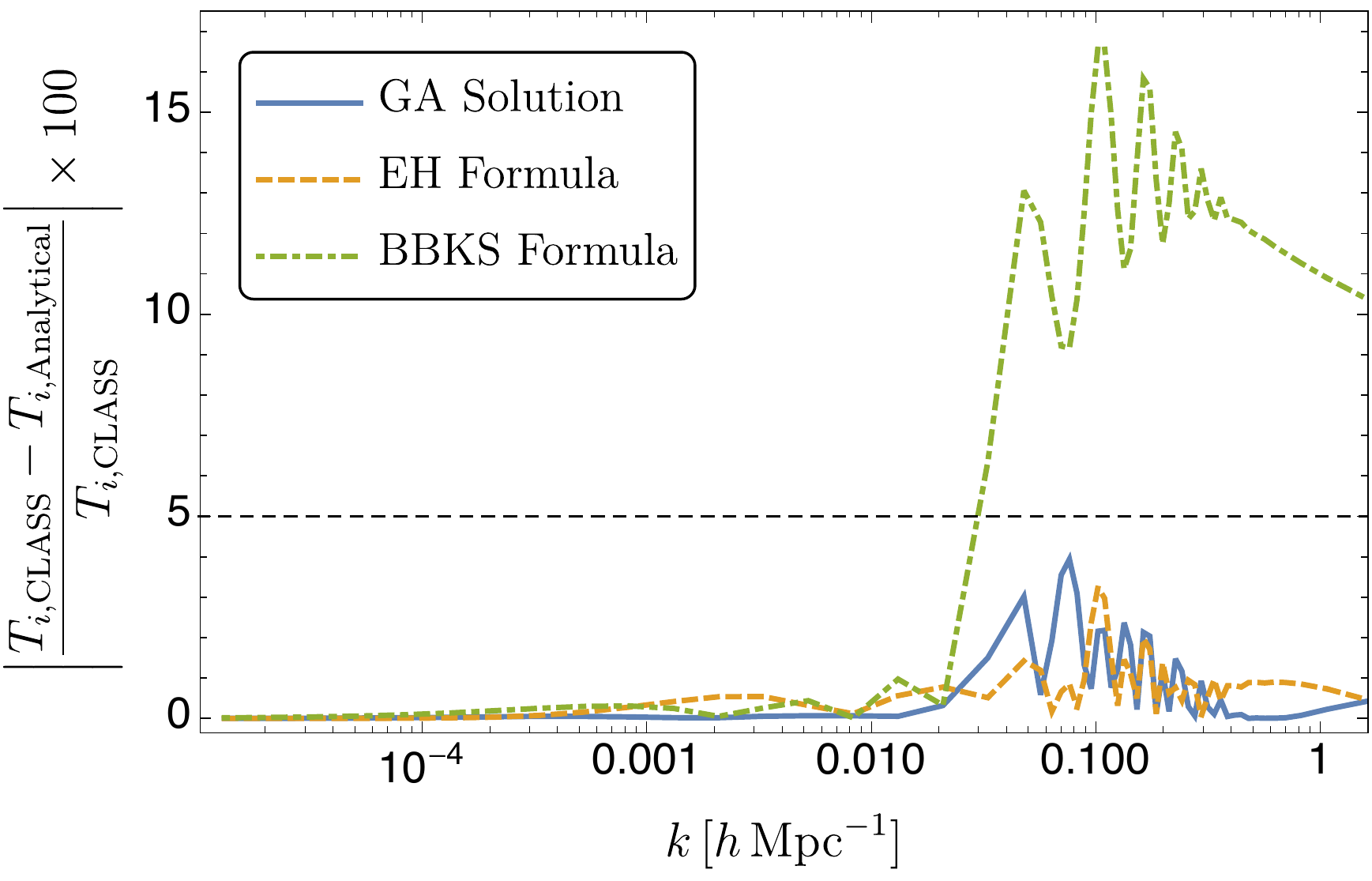}
\end{minipage}\qquad
\begin{minipage}[b]{.45\textwidth}
\includegraphics[width=\textwidth]{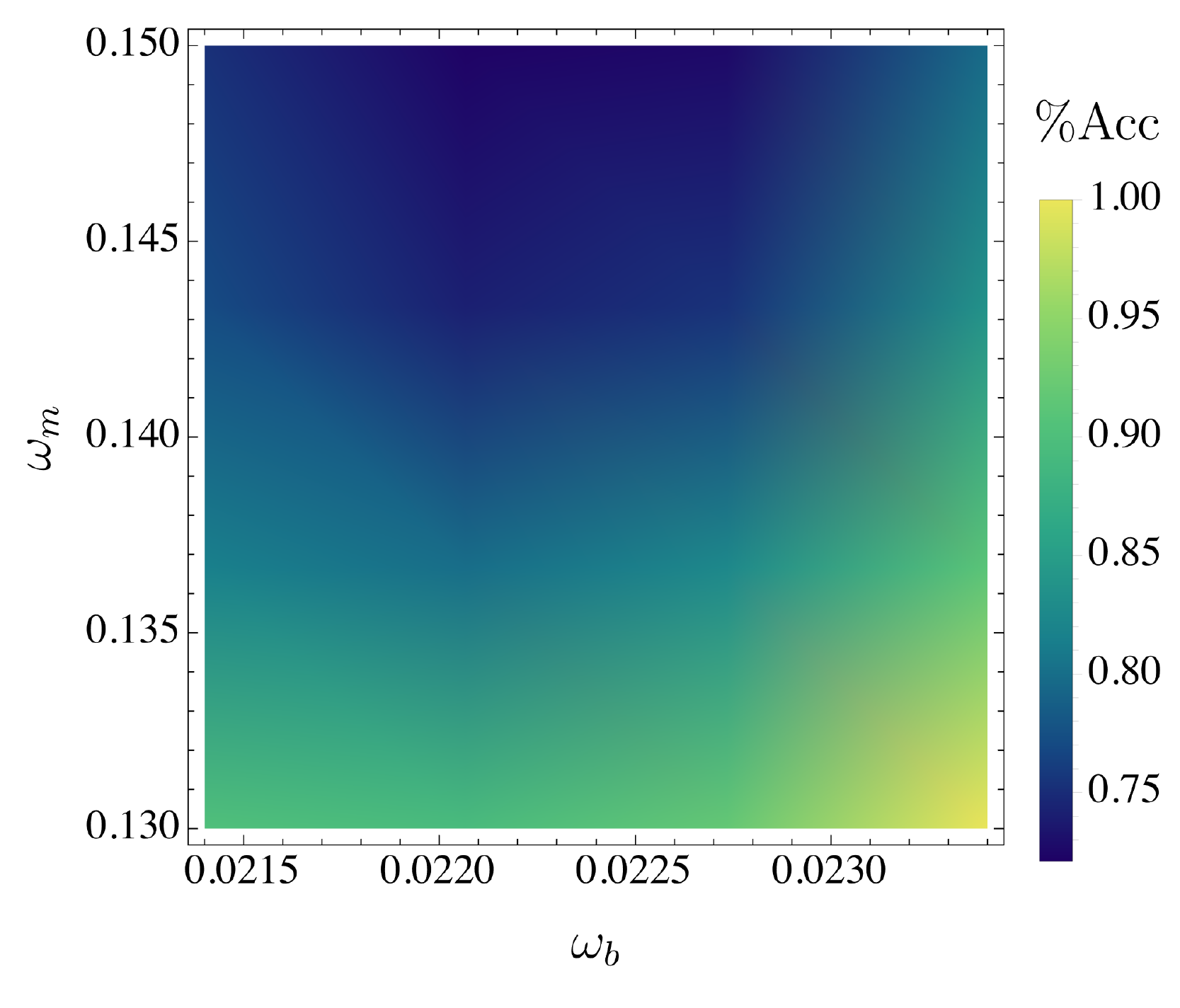}
\end{minipage}
\caption{(Left) Accuracy of the BBKS (green dot-dashed line), EH (mustard dashed line), and the GA (blue solid line) fitting formulae as a function of the scale $k$ and fixed $\{ \omega_m, \omega_b \}$. The parameters for these plots are: $\omega_b = 0.02273$, and $\omega_m = 0.1366$. Note that the accuracy of the three formulae notably decrease on the scales where the effects of baryons are most prominent on the cosmological evolution. However, the accuracy of the EH and GA formulae are below 5\% in this scale. (Right) Average accuracy of the GA fitting formula for all $k$ as a function of $\{\omega_m, \omega_b\}$. As it can be seen, the formula is up to $1\%$ accurate in overall.}
\label{Plot: Accuracy}
\end{figure*}

Now, we present our fitting formula when only baryons and matter are considered. Using the configuration of the GA as explained, the stochastic search ended up with the function
\begin{widetext}
\begin{align}
\label{Eq: TGA}
T_\text{GA}(k; \omega_b, \omega_m) &= \left[ 1 + 59.0998 \, x^{1.49177} + 4658.01 \, x^{4.02755} + 3170.79 \, x^{6.06} + 150.089 \, x^{7.28478} \right]^{-1/4},
\end{align}
\end{widetext}
whose goodness of fit, as measured by Eq.~\eqref{Eq: chi2 GA}, is $\%\text{Acc}_\text{GA} = 0.815012$. The accuracy of the BBKS formula in Eq. \eqref{Eq: BBKS} is $\%\text{Acc}_\text{BBKS} = 8.70957$, and for the EH formula is $\%\text{Acc}_\text{EH} = 0.777504$. We compile these results in Table \ref{Table: Accuracy}. In Fig.~\ref{Plot: Accuracy}, we compare the performance of the fitting formulae. As it can be seen in the left panel of this figure, the three formulae are very accurate on the large scales ($k < 0.01 h \, \text{Mpc}^{-1}$), while for smaller scales ($k > 0.01 h \, \text{Mpc}^{-1}$), where the effects of baryons are most prominent on the cosmological evolution, the accuracy of the three formulae decrease. However, we note that our GA formula and the EH formula can be accurate up to 5\%, while the BBKS formula is accurate up to 18\% in these scales. We would like to emphasize that our GA formula in Eq.~\eqref{Eq: TGA} is remarkably simpler than the EH formula which is described by around 30 expressions which occupy the whole Appendix~\ref{App: EH Formula}, while their difference in accumulative accuracy is just about 0.04. We also want to stress that, in principle, the GA could get even better results if some modifications are introduced. For instance, a larger grammar set and a more complex $f_\text{GA}$ function [instead of Eq.~\eqref{Eq: fGA}]. Nonetheless, our choices avoid over-fitting while yielding a smooth transfer function with the correct limits in $k$.

\subsection{Baryons, Matter and Massive Neutrinos}

\subsubsection{\textbf{Data}}

Apart of baryons and cold dark matter, massive neutrinos can contribute to the matter content once they are freely streaming. We can take into account the effects of massive neutrinos in the matter transfer function using \texttt{CLASS}. We compute $T(k)$ at redshift zero as a function of $\omega_b$, $\omega_m$ in the same ranges as in Sec. \ref{Sec: Baryons and Matter}, and $\omega_\nu \equiv 0.0107 (\sum_\nu m_\nu/ 1.0 \, \text{eV})$, assuming just one massive neutrino, and that the total mass of massive neutrinos is in the range $0.06 \, \text{eV} \leq \sum m_\nu \leq 0.12 \,\mathrm{eV}$. The lower bound of the later constraint corresponds to the minimum mass allowed from neutrino flavour oscillation experiments \cite{Lesgourgues:2013sjj}, and the upper bound is the maximum mass value allowed by Planck \cite{Planck:2018vyg}. In this case, we get a grid of $4 \times 4 \times 4$ data points for $\{ \omega_b, \omega_m , \omega_\nu \}$ and compute the gravitational potential for each triad. Again, for each considered cosmology (64 in total), \texttt{CLASS} retrieves $114$ points $\{k, \Phi \}$. Normalizing the potential to get the transfer function, our final data set is a table of dimensions $7296\times5$ with data points given as $\{k, \omega_b, \omega_m, \omega_\nu, T \}$. We use Eq. \eqref{Eq: chi2 GA} to quantify the goodness of fit of the analytical expressions, but now $N = 7296$ is the number of data points in our dataset, and $T_i$ is the transfer function evaluated at $k_i$, $\omega_{b_i}$, $\omega_{m_i}$, and $\omega_{\nu_i}$. 

\subsubsection{\textbf{Fitting Formula from the GA}}

\begin{figure*}
\centering
\begin{minipage}[b]{.45\textwidth}
\includegraphics[width=\textwidth]{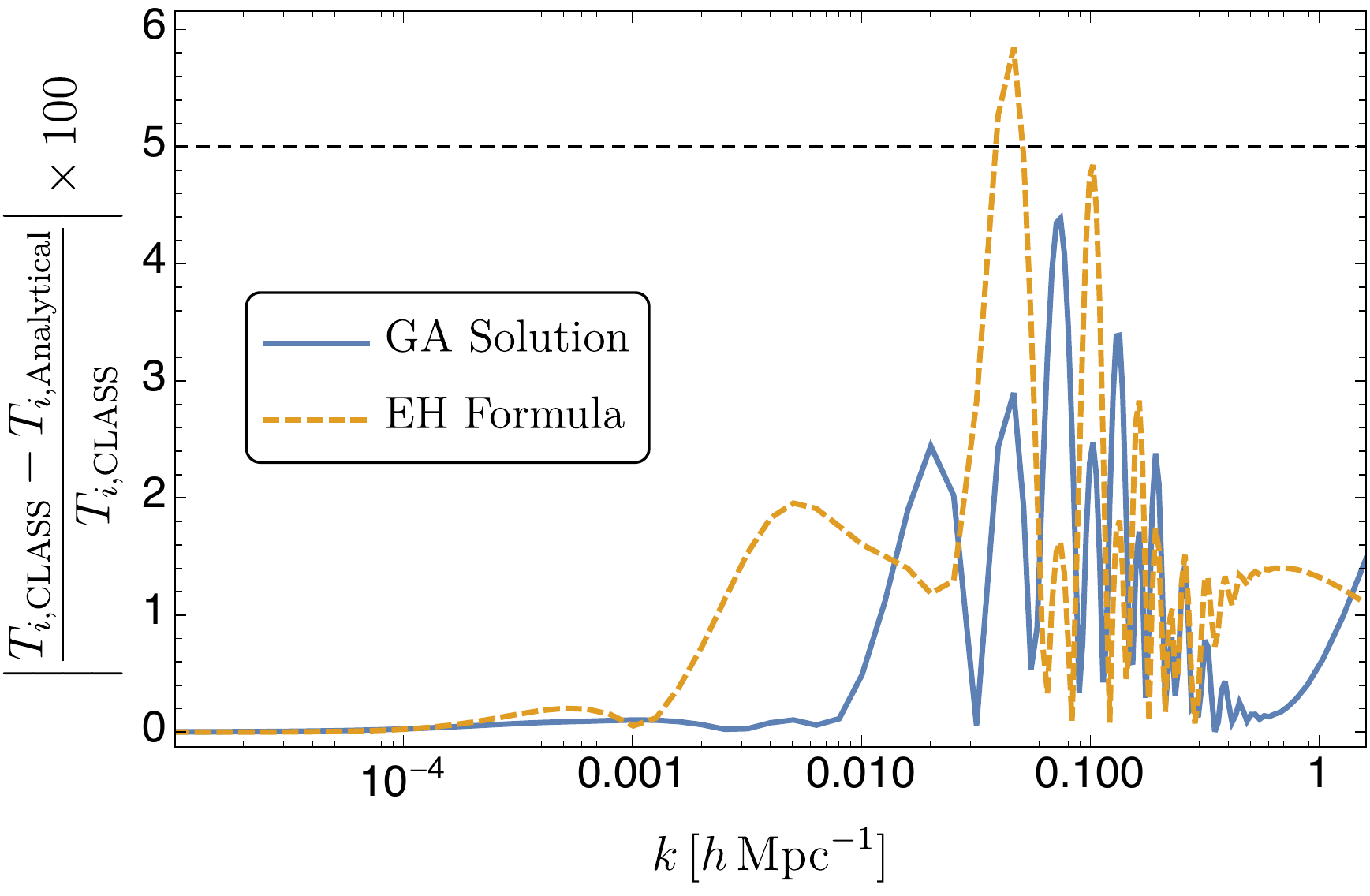}
\end{minipage}\qquad
\begin{minipage}[b]{.45\textwidth}
\includegraphics[width=\textwidth]{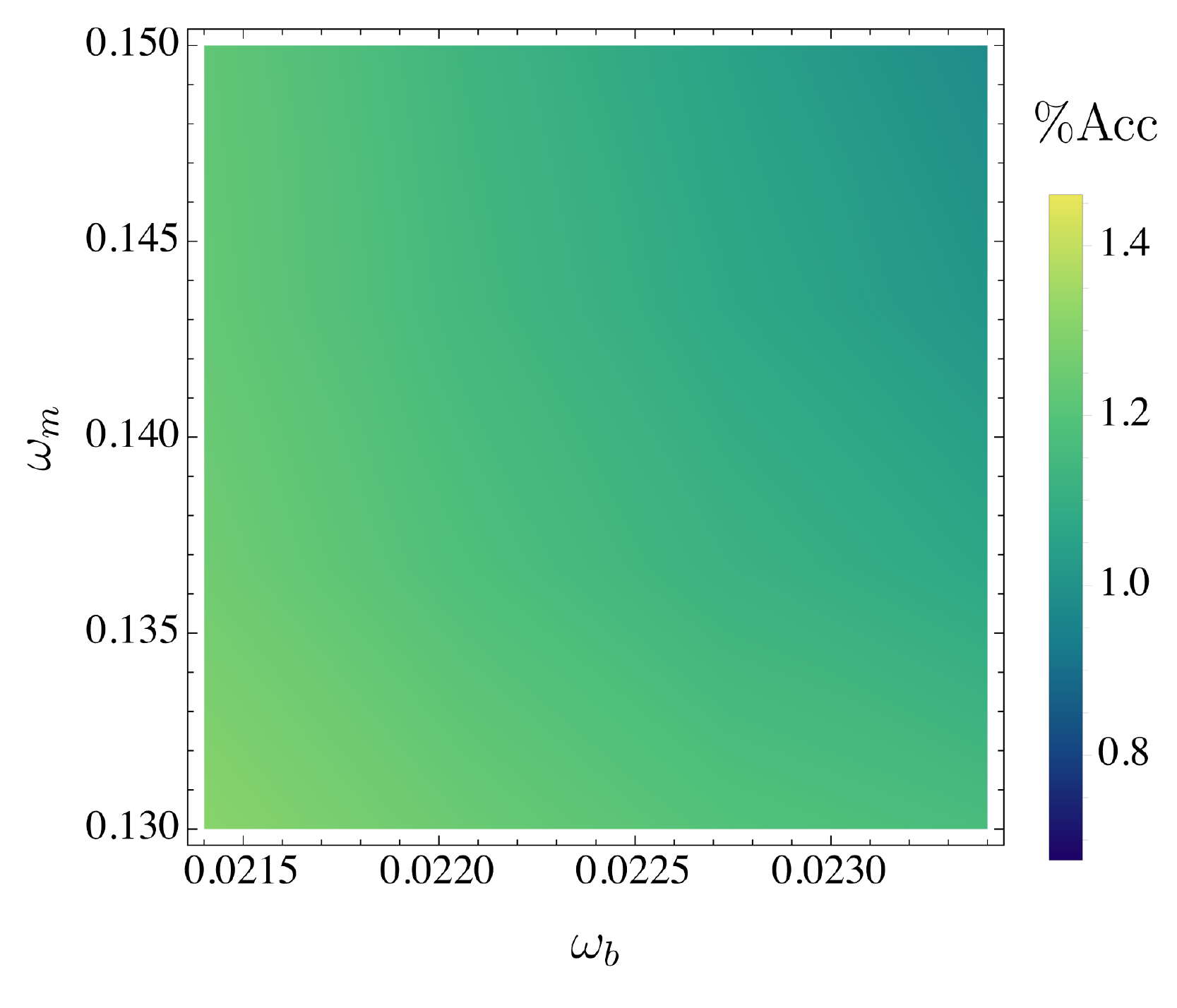}
\end{minipage}
\caption{(Left) Accuracy of the  EH (mustard dashed line), and the GA (blue solid line) fitting formulae as a function of the scale $k$ for fixed $\{\omega_m, \omega_b\, \omega_\nu \}$. The parameters for these plots are: $\omega_b = 0.02206$, $\omega_m = 0.1499$, and $\omega_\nu = 0.0008599$ which corresponds to a mass of $\sum_\nu m_\nu = 0.08 \, \text{eV}$. As can be seen, the accuracy of the GA formula is below 5\% on the whole range of $k$ considered here. (Right) Average accuracy of the GA fitting formula for all $k$ and $\omega_\nu$ as a function of $\{\omega_m, \omega_b\}$. As it can be seen, the formula is around $1\%$ accurate in overall.}
\label{Plot: Accuracy Neutrinos}
\end{figure*}

Based on the success of the GA in finding an accurate formula for $T(k)$ when neutrinos are massless, this time our GA is looking for a function of the form
\begin{equation}
\label{Eq: TGA 2}
T_\text{GA}(y) \equiv \left[ 1 + f_\text{GA}(y) \right]^{-1/4}, 
\end{equation}
where
\begin{equation}
\label{Eq: fGA 2}
y \equiv \frac{k \, \text{Mpc}}{\omega_m - \omega_b + \omega_\nu}, \quad f_\text{GA} \equiv \sum_{i = 1}^{4} a_i g_i \left( (b_i y
)^{c_i} \right).
\end{equation}
Now, we present our fitting formula which considers baryons, matter, and one massive neutrino:
\begin{widetext}
\begin{align}
\label{Eq: TGA Neutrino}
T_\text{GA}(k; \omega_b, \omega_m) &= \left[ 1 + 56.4933 \, y^{1.48261} + 3559.23 \, y^{3.76407} + 4982.44 \, y^{5.68246} + 374.167 \, y^{7.14558} \right]^{-1/4},
\end{align}
\end{widetext}
whose goodness of fit, as measured by Eq.~\eqref{Eq: chi2 GA}, is $\%\text{Acc}_\text{GA} = 0.993916$. The accuracy of the extended EH formula considering massive neutrinos is $\%\text{Acc}_\text{EH} = 1.23449$. We compile these results in Table~\ref{Table: Accuracy Neutrinos}. In Fig.~\ref{Plot: Accuracy Neutrinos}, we compare the performance of the fitting formulae. Similar to the last section, the formulae are more accurate on large scales ($k < 0.01 h \, \text{Mpc}^{-1}$) than on the smaller scales ($k > 0.01 h \, \text{Mpc}^{-1}$). Note in the left panel of this figure that the accuracy of the GA formula is below the $5\%$ in the whole range of $k$. Furthermore, the GA formula \eqref{Eq: TGA Neutrino} is significantly simpler than the EH formula, whose description requires around 30 expression which we give in full in Appendix~\ref{App: EH Formula Neutrinos}. Our result is a compelling analytical alternative to compute the matter transfer function.

\begin{table}[t!]
\setlength{\tabcolsep}{12pt}
\begin{tabular}{ccc}
\toprule 
 & Expression & $\%$ Accuracy \tabularnewline
\midrule
\midrule 
EH & Appendix~\ref{App: EH Formula Neutrinos} & 1.23449 \tabularnewline
GA & Eq. \eqref{Eq: TGA Neutrino} & 0.993916 
\tabularnewline
\bottomrule
\end{tabular}
\caption{Expression and accuracy of the fitting formulae for the matter transfer function as a function of $k$, $\omega_b$, $\omega_m$, and $\omega_\nu$. Although the accuracy of both formulations are similar, the simplicity of the GA function in Eq.~\eqref{Eq: TGA Neutrino} is a major improvement over the EH formula, which is fully displayed in Appendix~\ref{App: EH Formula Neutrinos}.}
\label{Table: Accuracy Neutrinos}
\end{table}

\section{Conclusions} 
\label{Sec: Conclusions}

Genetic Algorithms (GAs), a machine learning technique, have been previously used  to obtain improved formulation of cosmological quantities of interest such as the redshift at recombination and the sound horizon at drag epoch \cite{Aizpuru:2021vhd}, and also to perform consistency tests to probe for deviations from $\Lambda$CDM \cite{Renzi:2020bvl, EUCLID:2020syl, Arjona:2020axn, arjona:2021mzf, Euclid:2021frk,  Arjona:2021hmg, Alestas:2022gcg}. Here we used the GA to find analytical alternatives to existing fitting functions for the matter transfer function, such as the BBKS and the EH formulae. 

When the transfer function depends only on $\omega_b$ and $\omega_m$, the GA finds a very simple fitting formula for $T(k)$ [see Eq.~\eqref{Eq: TGA}], which is as accurate as the EH formula, while being significantly shorter (see Appendix~\ref{App: EH Formula}). In a more realistic scenario, at least one neutrino should be massive. In this case, our GA finds a fitting function which is both more accurate and notably simpler [see Eq.~\eqref{Eq: TGA Neutrino}] than the EH formula in Appendix~\ref{App: EH Formula Neutrinos}. The simplicity of the GA fitting formulae and their accuracy represent a major improvement over other existing analytical formulations of the matter transfer function which are extensively used in the literature \cite{Boyanovsky:2008he, Dvornik:2022xap, Schoneberg:2022ggi}. Therefore, the GA formulae in Eqs.~\eqref{Eq: TGA} and \eqref{Eq: TGA Neutrino} are compelling semi-analytic  alternatives to easily and accurately compute the matter power spectrum.

\section*{Acknowledgements}
The authors would like to thank R.~Arjona for useful discussions at a very early stage of this project, and also to Jose Palacios for carefully reading the manuscript and providing useful comments. BOQ would like to express his gratitude to the  Instituto de F\'isica Téorica UAM-CSIC, in Universidad Autonóma de Madrid, for the hospitality and kind support during all the  stages of this work. BOQ is also supported by Patrimonio Aut\'onomo - Fondo Nacional de Financiamiento para la Ciencia, la Tecnolog\'ia y la Innovaci\'on Francisco Jos\'e de Caldas (MINCIENCIAS - COLOMBIA) Grant No. 110685269447 RC-80740-465-2020, projects 69723 and 69553. WC acknowledges financial support from the S\~ao Paulo Research Foundation (FAPESP) through grant \#2021/10290-2. This research was supported by resources supplied by the Center for Scientific Computing (NCC/GridUNESP) of the S\~ao Paulo State University (UNESP). SN acknowledges support from the research projects PGC2018-094773-B-C32 and PID2021-123012NB-C43, and by the Spanish Research Agency (Agencia Estatal de Investigaci\'on) through the Grant IFT Centro de Excelencia Severo Ochoa No CEX2020-001007-S, funded by MCIN/AEI/10.13039/501100011033.

\section*{Numerical Codes}

The genetic algorithm code used here can be found in the GitHub repository \href{https://github.com/BayronO/Transfer-Function-GA}{https://github.com/BayronO/Transfer-Function-GA} of BOQ. This code is based on the GA code by SN which can be found at \href{https://github.com/snesseris/Genetic-Algorithms}{https://github.com/snesseris/Genetic-Algorithms}.

\appendix

\section{EH Formula}
\label{App: EH Formula}

The transfer function given by Eisenstein and Hu \cite{Eisenstein:1997ik} has the following form:
\begin{equation}
T(k) = \frac{\Omega_b}{\Omega_0} T_b(k) + \frac{\Omega_c}{\Omega_0} T_c(k),
\end{equation}
where $\Omega_0 = \Omega_b + \Omega_c$. The terms involved in this formula are the following:
\begin{equation}
T_b = \Bigg[\frac{\tilde{T}_0(k; 1,1)}{1 + \left( ks/ 5.2 \right)^2} + 
\frac{\alpha_b}{1 + \left(\beta_b / ks \right)^3} e^{-\left( \frac{k}{k_{\mathrm{Silk}}} \right)^{1.4}}\Bigg] j_0(k \tilde{s}),
\end{equation}
\begin{equation}
T_c = f \tilde{T}_0(k, 1, \beta_c) + (1 - f) \tilde{T}_0(k, \alpha_c, \beta_c),
\end{equation}
\begin{equation}
\tilde{T}_0(k, \alpha_c, \beta_c) = \frac{ \ln(e + 1.8 \beta_c q)}{\ln(e + 1.8 \beta_c q) + C q^2},
\end{equation}
\begin{align}
f &= \frac{1}{1+(ks/5.4)^4}, \\
C &= \frac{14.2}{\alpha_c} + \frac{386}{1 + 69.9 q^{1.08}}, \\
q &= \frac{k}{13.41 k_{\mathrm{eq}}}, \\
R &\equiv 3\rho_b/4\rho_\gamma = 
31.5 \omega_b \Theta_{2.7}^{-4} (z/10^3)^{-1},
\end{align}
\begin{align}
k_{\mathrm{Silk}} &= 1.6 \omega_b^{0.52} \omega_0^{0.73} \left[1 + \left( 10.4 \, \omega_0 \right)^{-0.95} \right] \,  \mathrm{Mpc}^{-1}, \\
k_{\mathrm{eq}} &= 7.46 \times 10^{-2} \omega_0 \Theta_{2.7}^{-2} \, \mathrm{Mpc}^{-1},
\end{align}
\begin{align}
\alpha_b &= 2.07 k_{\mathrm{eq}} s (1 + R_d)^{-3/4} G \left( \frac{1 + z_{\mathrm{eq}}}{1 + z_d} \right), \\
\beta_b &= 0.5 + \frac{\Omega_b}{\Omega_0} + \left( 3 - 2 \frac{\Omega_b}{\Omega_0} \right) \sqrt{\left( 17.2 \, \omega_0 \right)^2 + 1}, \\
\beta_{\mathrm{node}} &= 8.41 \omega_0^{0.435},
\end{align}
\begin{align}
s &= \frac{2}{3 k_{\mathrm{eq}}} \sqrt{\frac{6}{R_{\mathrm{eq}}}} \ln \frac{\sqrt{1 + R_d} + \sqrt{ R_d + R_{\mathrm{eq}}}}{1 + \sqrt{R_{\mathrm{eq}}}}, \\
\tilde{s} &= s \left[1 + \left( \frac{\beta_{\mathrm{node}}}{k s} \right)^3 \right]^{-1/3},
\end{align}
\begin{align}
G(y) &= - 6 y \sqrt{1 + y} \nonumber \\
 &+ y (2 + 3y) \ln \left(\frac{
\sqrt{1 + y} + 1}{\sqrt{1 + y} -1} \right), \\
y &\equiv \frac{1 + z_\text{eq}}{1 + z},
\end{align}
\begin{align}
\alpha_c &= a_1^{-\Omega_b/\Omega_0} a_2^{-(\Omega_b/\Omega_0)^3}, \\
a_1 &= (46.9 \omega_0)^{0.670} [ 1 + (32.1 \omega_0)^{-0.532}], \\
a_2 &= (12.0 \omega_0)^{0.424} [1 + (45.0 \omega_0 )^{-0.582}],
\end{align}
\begin{align}
\beta_c^{-1} &= 1 + b_1 [(\Omega_c/\Omega_0)^{b_2} - 1], \\
b_1 &= 0.944 [ 1 + (458 \omega_0)^{-0.708} ]^{-1}, \\
b_2 &= (0.395 \omega_0)^{-0.0266}.
\end{align}
\begin{align}
z_{\mathrm{eq}} &= 2.50 \times 10^4 \omega_0 \Theta_{2.7}^{-4}, \\
z_d &= 1291 \frac{\omega_0^{0.251}
}{1 + 0.659 \omega_0^{0.828}} \left[ 1 + b_{1, z} \omega_b^{b_{2, z}} \right], \\
b_{1, z} &= 0.313 \omega_0^{-0.419} \left[ 1 + 0.607 
\omega_0^{0.674} \right], \\
b_{2, z} &= 0.238 \omega_0^{0.223},
\end{align}
where we have defined $\omega_0 = (\Omega_c + \Omega_b) h^2$, and $T_\text{CMB} \equiv 2.7 \Theta_{2.7} \, \text{K}$, $R_d \equiv R(z_d)$ and $R_\text{eq} \equiv R(z_\text{eq})$

\section{EH Formula with Massive Neutrinos}
\label{App: EH Formula Neutrinos}

The transfer function given by Eisenstein and Hu considering massive neutrinos has the following form \cite{Eisenstein:1997jh}:
\begin{equation}
T_{cb\nu}(k, z) = T_\text{master}(k) \frac{D_{cb\nu}(k, z)}{D_1(z)},
\end{equation}
The terms involved in this formula are the following:
\begin{align}
T_\text{master} (k) &= T_\text{sup} (k) B(k), \\
D_{cb\nu}(z, q) &= \left[f_{cb}^{0.7/p_{cb}} + \left(\frac{D_1(z)}{1 + y_{\text{fs}}(q; f_\nu)}\right)^{0.7} \right]^{p_{cb}/0.7} \nonumber \\
&\times D_1(z)^{1 - p_{cb}}, 
\end{align}
\begin{align}
D_1(z) &= \left(\frac{1 + z_{eq}}{1+z}\right)\frac{5\Omega(z)}{2}\Big\{\Omega(z)^{4/7} \\
& - \Omega_\Lambda(z) + \left[1 + \frac{\Omega(z)}{2} \right]\left[1 + \frac{\Omega_\Lambda(z)}{70} \right] \Big\}^{-1}, \nonumber \\
\Omega(z) &= \Omega_0 (1+z)^3 g^{-2}(z), \\
\Omega_\Lambda(z) &= \Omega_\Lambda g^{-2}(z), \\
g^2(z) &= (1-\Omega_0-\Omega_\Lambda)(1+z)^2 +\Omega_0 (1+z)^3 \nonumber \\
& +\Omega_\Lambda,
\end{align}
\begin{align}
T_\text{sup} (k) &= \frac{L}{L + C q^2_\text{eff}}, \\ 
L &= \ln (e +1.84 \beta_c \sqrt{\alpha_\nu} q_{\text{eff}}), \\
C &= 14.4 + \frac{325}{1 + 60.5 q_{\text{eff}}^{1.08}}, \\
\beta_c &= (1 - 0.949 f_{\nu b})^{-1},
\end{align}
\begin{align}
y_{\rm fs}(q) &= 17.2 f_\nu(1 + 0.488f_\nu^{-7/6})(q N_\nu / f_\nu)^2, \\
q &= \frac{k}{\text{Mpc}^{-1}} \Theta_{2.7}^2 \omega_0^{-1},
\end{align}
\begin{align}
B(k) &= 1+\frac{1.24f_\nu^{0.64}N_\nu^{0.3+0.6f_\nu}}
{q_\nu^{-1.6}+q_\nu^{0.8}},\\
q_\nu &=
3.92 q\sqrt{\frac{N_\nu}{f_\nu}}
\end{align}
\begin{align}
f_c &= \frac{\Omega_b}{\Omega_m + \Omega_\nu}, \quad f_\nu = \frac{\Omega_\nu}{\Omega_m + \Omega_\nu}, 
\nonumber \\
f_{cb} &= \frac{\Omega_m}{\Omega_m + \Omega_\nu}, \quad f_{\nu b} = \frac{\Omega_b + \Omega_\nu}{\Omega_m + \Omega_\nu},
\end{align}
\begin{align}
q_\text{eff} &= \frac{k \Theta_{2.7}^2}{\Gamma_{\text{eff}} \, \text{Mpc}^{-1}}, \\
\Gamma_{\text{eff}} &= \omega_0 \left[\sqrt{\alpha_\nu} + 
\frac{1 - \sqrt{\alpha_\nu}}{1 + (0.43 k s)^4} \right],
\end{align}
\begin{align}
\alpha_\nu &= \frac{f_c}{f_{cb}} \frac{5 - 2 (p_c + p_{cb})}{5 - 4 p_{cb}} \\
 &\times \frac{1 - 0.553 f_{\nu b} + 0.126 f_{\nu b}^3}{1 - 0.193 \sqrt{f_\nu N_\nu} + 0.169 f_\nu N_\nu^{0.2}}(1 + y_d)^{p_{cb} - p_{c}} \nonumber \\
 &\times \Big[1 + \frac{p_{c} - p_{cb}}{2} \left(1 + \frac{1}{(3 - 4 p_c)(7 - 4 p_{cb})} \right) \nonumber \\
 &\times (1 + y_d)^{-1}\Big], \\
p_c &\equiv \frac{1}{4} \left[ 5 - \sqrt{1 + 24 f_c} \right], \\
p_{cb} &\equiv \frac{1}{4} \left[ 5 - \sqrt{1 + 24 f_{cb}} \right],
\end{align}
\begin{align}
z_{\mathrm{eq}} &= 2.50 \times 10^4 \omega_0 \Theta_{2.7}^{-4}, \\
z_d &= 1291 \frac{\omega_0^{0.251}
}{1 + 0.659 \omega_0^{0.828}} \left[ 1 + b_{1, z} \omega_b^{b_{2, z}} \right], \\
b_{1} &= 0.313 \omega_0^{-0.419} \left[ 1 + 0.607 
\omega_0^{0.674} \right], \\
b_{2} &= 0.238 \omega_0^{0.223}, \\
y &\equiv \frac{1 + z_\text{eq}}{1 + z}, \\
s &= \frac{44.5 \ln (9.83 \omega_0)}{\sqrt{1 + 10 \omega_b^{3/4}}} \, \text{Mpc},
\end{align}
where $N_\nu$ is the number of massive neutrinos, $\Omega_\Lambda$ is the density parameter of cosmological constant, and we have redefine some terms like $\Omega_0 = \Omega_c + \Omega_b + \Omega_\nu$, and $\omega_0 = \Omega_0 h^2$.

\bibliographystyle{utcaps} 
\bibliography{Bibli.bib}

\end{document}